\documentclass[preprintnumbers,floatfix,letterpaper,superscriptaddress,nofootinbib,twocolumn]{revtex4-2}
\pdfoutput=1
\usepackage{amsmath}
\usepackage{amssymb}
\usepackage{bbm}
\usepackage{subfigure}
\usepackage{hyperref}
\usepackage{url}
\usepackage[dvipdfmx]{graphicx}
\usepackage{xcolor}
\usepackage{color}
\usepackage{mathrsfs}
\usepackage{amsfonts}
\usepackage{latexsym}
\usepackage{ragged2e}
\usepackage{epsfig}
\usepackage{textcomp}
\usepackage{phaistos}
\usepackage[title]{appendix}
\usepackage{chngcntr}
\usepackage{microtype}

\makeatletter
\renewcommand\@makefnmark{\hbox{\@textsuperscript{\normalfont\color{purple}\@thefnmark}}}
\renewcommand\@makefntext[1]{%
  \parindent 1em\noindent
            \hb@xt@1.8em{%
                \hss\@textsuperscript{\normalfont\@thefnmark}}#1}
\makeatother

\usepackage{caption}
\DeclareCaptionJustification{justified}{\leftskip=0pt \rightskip=0pt \parfillskip=0pt plus 1fil}
\graphicspath{{./Figures/}}

\def\beq{\begin{equation}}
\def\eeq{\end{equation}}

\def\mathbb{\Bbb}

\definecolor{vividviolet}{rgb}{0.62, 0.0, 1.0}
\definecolor{amaranth}{rgb}{0.9, 0.17, 0.31}
\definecolor{palatinateblue}{rgb}{0.15, 0.23, 0.89}
\definecolor{brightpink}{rgb}{1.0, 0.0, 0.5}
\definecolor{cornflowerblue}{rgb}{0.39, 0.58, 0.93}
\definecolor{deepcarminepink}{rgb}{0.94, 0.19, 0.22}
\definecolor{radicalred}{rgb}{1.0, 0.21, 0.37}
\colorlet{RED}{red}

\hypersetup{linktoc=all,
    colorlinks, linkcolor={palatinateblue},
    citecolor={brightpink}, urlcolor={amaranth}
}
\newcommand{\changeurlcolor}[1]{\hypersetup{urlcolor=#1}}

\renewcommand{\d}[1]{\ensuremath{\operatorname{d}\!{#1}}}

\renewcommand{\d}[1]{\ensuremath{\operatorname{d}\!{#1}}}
\makeatletter
\def\@fnsymbol#1{\ensuremath{\ifcase#1\or $\textleaf$ \or $\PHplaneTree$
\else\@ctrerr\fi}}%

\makeatother
\def\sideremark#1{\ifvmode\leavevmode\fi\vadjust{\vbox to0pt{\vss
 \hbox to 0pt{\hskip\hsize\hskip1em
 \vbox{\hsize1.5cm\tiny\raggedright\pretolerance10000
 \noindent #1\hfill}\hss}\vbox to8pt{\vfil}\vss}}}

\begin{document}
\title{On the Counter-Rotation of Closed Timelike Curves}

\author{Yuanyuan Duan}
\affiliation{Center for Gravitation and Cosmology, College of Physical Science and Technology, Yangzhou University, \\180 Siwangting Road, Yangzhou City, Jiangsu Province  225002, China}

\author{Fangxun Liu}
\affiliation{Center for Gravitation and Cosmology, College of Physical Science and Technology, Yangzhou University, \\180 Siwangting Road, Yangzhou City, Jiangsu Province  225002, China}

\author{Yu Wang}
\affiliation{Center for Gravitation and Cosmology, College of Physical Science and Technology, Yangzhou University, \\180 Siwangting Road, Yangzhou City, Jiangsu Province  225002, China}

\author{Yen Chin \surname{Ong}}
\email{ycong@yzu.edu.cn}
\affiliation{Center for Gravitation and Cosmology, College of Physical Science and Technology, Yangzhou University, \\180 Siwangting Road, Yangzhou City, Jiangsu Province  225002, China}
\affiliation{School of Aeronautics and Astronautics, Shanghai Jiao Tong University, Shanghai 200240, China}

\begin{abstract}
While it is tempting to think of closed timelike curves (CTCs) around rotating bodies such as a black hole  as being ``caused'' by the rotation of the source, Andr\'eka et al. pointed out that the underlying physics is not as straightforward since such CTCs are ``counter-rotating'', i.e., the time orientation (the opening of the local light cones) of the CTCs is opposite to the direction in which the singularity or the ergosphere rotates. It was also suggested that this is a generic phenomenon that calls for a deeper intuitive physical understanding. In this short note we point out -- with Kerr-Taub-NUT as an example -- that CTCs are counter-rotating with respect to the \emph{local} angular velocity of the spacetime, which makes a physical interpretation of CTCs being ``caused'' by a rotating source even more problematic. 
\end{abstract}

\maketitle

\section{The ``Counter-Rotation'' of Closed Timelike Curves}\label{intro}

It is well known that closed timelike curves occur in a wide variety of rotating spacetime geometries at the formal mathematical level. Though actual physical time travel does not seem feasible, it is nevertheless interesting to study these oddities that could lead to some deeper insights about spacetimes and gravity. Even if time travel is not permitted, understanding how chronology protection \cite{chrono} works may teach us more about quantum gravity \cite{0204022}. See also \cite{2101.08592} for a recent historical survey of CTC physics. 

In \cite{0708.2324}, Andr\'eka et al. discussed one interesting property of CTCs: their time orientation is opposite to the direction of the rotating body. This is true in the geometries of Kerr-Newman black hole, the Kerr black hole (for which CTC only exists for $r<0$), the Tipler-van Stockum cylinder \cite{T1, T2, V}, the G\"odel universe \cite{Godel, 1303.4651}, as well as Gott's cosmic strings based CTCs \cite{Gott}. Thus, it is natural for Andr\'eka et al. to conjecture that this counter-rotating phenomenon is a generic one, which could provide some hints to understand the formation of CTCs. The fact that these CTCs are counter-rotating means that it is rather problematic (or at least not obvious) to think of CTCs as being the product of the extreme rotation of the gravitational source, similar to frame dragging. In fact, one of the questions posed in \cite{0708.2324} is the following:
\emph{
``How important is it for the CTCs to counter-rotate against the rotational sense of the gravitating matter which brings about the
CTCs? In particular, is there any example of a spacetime where the CTCs are generated by rotating matter and there is no counter-rotation effect?''}

In this short note, we begin by a short review on the method of determining the orientation of CTCs in Sec.(\ref{II}), from which we can see why such a counter-rotation arises mathematically. 
We emphasize that the mathematics is clear, the problem is at the level of physical understanding: are CTCs caused by the angular momentum/velocity of the source, and if so how are they formed?
Ultimately it is the dynamical process that is most interesting \footnote{If one can in principle design a time machine, its aim would be to create CTCs from a spacetime that is devoid of them in the beginning. Even understanding the proper formulation of this problem is not easy, see \cite{Earman} for a recent attempt.}, but the first step would be to try to understand in more details the relationship between the angular momentum of the source and that of the orientation of the CTCs. We then show that the CTCs near the north pole of the Kerr-Taub-NUT (Newman-Unti-Tambourino) black hole co-rotate with respect to the globally defined nonzero angular momentum (which is in the same direction as the angular velocity of the event horizon), but nevertheless, are counter-rotating with respect to the \emph{local} angular velocity. Finally we discuss the implication of our results. We will work in the units $G=c=1$.

\section{Closed Timelike Curves and Their Orientation}\label{II}

Most rotating spacetimes can be put into an axial-symmetric cylindrical-type coordinates, which include the Boyer-Lindquist coordinates for Kerr and similar black holes. The co-latitudal angular coordinate $\phi$ is associated with the vector field ${\partial}/{\partial \phi}$, the integral curves of which are closed. The location of closed timelike curves are thus given by $g_{\phi\phi}<0$ (CTCs are nevertheless independent of coordinates). The angular velocity experienced by a test particle in a rotating spacetime is given by $\Omega:=-g_{t\phi}/g_{\phi\phi}$. When evaluated on the event horizon, this gives the ``angular velocity of the horizon'', often denoted by $\Omega_+$. For the Kerr black hole of mass $M$ and angular momentum $J$, for example, $\Omega_+ \propto a=J/M$, so that the rotation parameter $a$ gives the same direction of rotation as $\Omega_+$. In a more complicated spacetime, $a$ need not have the same sign as $\Omega_+=\Omega_+(r,\theta)$ for all values of $r$ and $\theta$. 

The orientation of the CTC is best understood locally by looking at the light cones adapted to the aforementioned ``cylindrical coordinates''. To introduce this idea, let us first consider a 2-dimensional flat spacetime and the two well-known metrics 
\begin{itemize}
\item[(1)] Minkowski metric: $\d s^2 = -\d t^2 + \d x^2$,
\item[(2)] Rindler metric: $\d s^2 = -X^2\d T^2 + \d X^2$.
\end{itemize}
We can re-write these in matrix notation as:
\begin{itemize}
\item[(1)] Minkowski metric: \\$\d s^2 = (\d t ~\d x)\begin{pmatrix} -1 & 0 \\ 0 & 1  \end{pmatrix} \begin{pmatrix} \d t \\ \d x\end{pmatrix} \equiv v^T\eta v$,
\item[(2)] Rindler metric: \\$\d s^2 = (\d T ~ \d X)\begin{pmatrix} -X^2 & 0 \\ 0 & 1 \end{pmatrix} \begin{pmatrix} \d T \\ \d X\end{pmatrix} \equiv u^Tgu$.
\end{itemize}
We note that if there exists a transformation $e$ such that $e^Tge=\eta$, then the Minkowski light cone ($\d s^2=0$) satisfies $v^Te^Tgev=0$, i.e., $(ev)^Tg(ev)=0$. The plan is to find a transformation matrix $e$ such that $ev=u$. Such $e$ is not unique due to Lorentz symmetry. One obvious choice is to set
\begin{equation}\notag
e =\begin{pmatrix} 
X^{-1} & 0 \\ 0 & 1
\end{pmatrix}.
\end{equation}

Now, consider a light cone in Minkowski spacetime and label the two null directions by  $(1, 1)^T$ and $(1,-1)^T$ respectively. Consider the mapping of these vectors under the transformation matrix $e$. We see that they get mapped as follows:
\begin{equation}\notag
\begin{pmatrix}
X^{-1} & 0 \\ 0 & 1
\end{pmatrix}
\begin{pmatrix}
1 \\ 1
\end{pmatrix}
=
\begin{pmatrix}
X^{-1} \\ 1
\end{pmatrix}, ~~
\begin{pmatrix}
X^{-1} & 0 \\ 0 & 1
\end{pmatrix}
\begin{pmatrix}
1 \\ -1
\end{pmatrix}
=
\begin{pmatrix}
X^{-1} \\ -1
\end{pmatrix}.
\end{equation}
These yield the light cones in the Rindler coordinates.

For axial-symmetric spacetimes, we focus on the ($t$-$\phi$)-plane:
\begin{equation}\notag
g[(t,\phi)] = \begin{pmatrix} \d t & r \d\phi \end{pmatrix} \begin{pmatrix} g_{tt} & {g_{t\phi}}/{r} \\  {g_{t\phi}}/{r} &  {g_{\phi\phi}}/{r^2} \end{pmatrix} \begin{pmatrix} \d t \\ r \d\phi \end{pmatrix}.
\end{equation}
As above, we would like to have a matrix $e$ such that
\begin{equation}\notag
e^T g e =\eta.
\end{equation}
By local Lorentz freedom one could set one the elements in $e$ to be zero:
\begin{equation}\notag
\begin{pmatrix} \alpha & 0 \\ \beta & \delta \end{pmatrix} 
\begin{pmatrix} g_{tt} & g_{t\phi}/r \\ g_{t\phi}/r & g_{\phi\phi}/r^2 \end{pmatrix} 
\begin{pmatrix}\alpha & \beta \\ 0 & \delta \end{pmatrix} = \begin{pmatrix} -1 & 0 \\ 0 & 1 \end{pmatrix}.
\end{equation}
We can then solve for $\alpha, \beta$ and $\delta$:
\begin{equation}\notag
\alpha = \frac{1}{\sqrt{-g_{tt}}} > 0; ~~ \beta = \frac{\text{sgn}(g_{t\phi})}{\sqrt{1-\frac{g_{tt}g_{\phi\phi}}{g_{t\phi}^2}}}\alpha, ~~ \delta = \frac{-r g_{tt}}{g_{t\phi}}\beta > 0.
\end{equation}

Here we imposed the conditions that $\alpha$ and $\delta$ are positive; this means that the light cones are not ``reflected'' under the transformation $e$. From the expression of $\beta$, this also implies that we must have $\text{sgn}(g_{t\phi}) = \text{sgn}(\beta)$. 
Under the transformation $e$, the light cones $(1, 1)^T$ and $(1,-1)^T$ are mapped into:
\begin{equation}\notag
\begin{pmatrix}
\alpha & \beta \\ 0 & \delta 
\end{pmatrix}
\begin{pmatrix}
1 \\ -1
\end{pmatrix}
=
\begin{pmatrix}
\alpha - \beta \\ -\delta 
\end{pmatrix}, ~~
\begin{pmatrix}
\alpha & \beta \\ 0 & \delta 
\end{pmatrix}
\begin{pmatrix}
1 \\ 1
\end{pmatrix}
=
\begin{pmatrix}
\alpha + \beta \\ \delta
\end{pmatrix},
\end{equation}
respectively.

\begin{figure}[htbp]
\centering
\includegraphics[width=1.05\columnwidth,keepaspectratio]{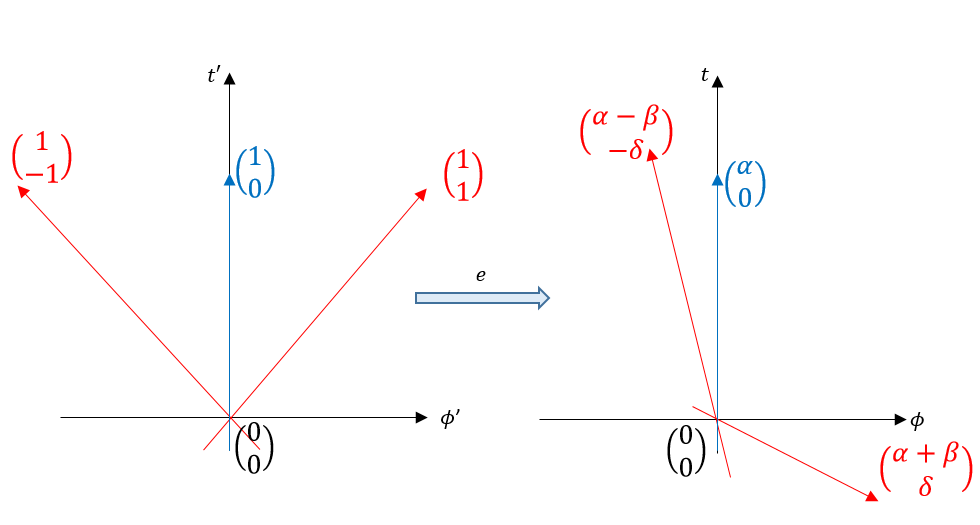}
\caption{\label{fig:1} Schematic illustation (not to scale) of the transformation $e$ that maps the Minkowski light cone to the light cone in rotating CTC spacetime. It is possible -- as is illustrated here -- that $\alpha + \beta$ is tilted so much forward that it tipped below the spatial axis, giving rise to a CTC. In this case $\alpha + \beta < 0$, and the CTC is counter-rotating with respect to $\Omega < 0$ (i.e. in the decreasing $\phi$ direction) and $g_{t\phi}<0$.}
\end{figure}

This simple linear algebra approach thus reveals two possibilities regarding the tilt of the light cones:
\begin{itemize}
\item[(i)]\begin{equation}\notag
g_{t\phi} > 0, ~~\beta > 0 \Longrightarrow \alpha - \beta < \alpha + \beta,
\end{equation}
\item[(ii)] \begin{equation}\notag
g_{t\phi} < 0, ~~\beta < 0 \Longrightarrow \alpha - \beta > \alpha + \beta.
\end{equation}
\end{itemize}
Either $\alpha+\beta$ or $\alpha-\beta$ must be negative for CTC to exist (i.e. the light cone opens wide enough to encompass the $\phi$-axis, so that $\partial/\partial\phi$ is timelike).
The first possibility corresponds to light cones that tilts ``backward'' with respect to the increasing $\phi$ direction.
This is in the direction of $\Omega$, if $\Omega$ is positive.
That is, the CTC is counter-rotating. This is the only possibility because CTC has to satisfy $g_{\phi\phi}<0$, which implies that $\text{sgn}(\Omega)=\text{sgn}(g_{t\phi})$.
Likewise, the second possibility corresponds to a counter-rotating CTC for $\Omega<0$. See Fig.(\ref{fig:1}) for an illustration. This is why a co-rotating CTC does not exist. 

To conclude, a negative $g_{t\phi}$ means that we have a ``forward tilt'' in the increasing $\phi$ direction, but
this only happens when $\Omega<0$; likewise $g_{\phi\phi}>0$ corresponds to a ``backward tilt'' and occurs only when $\Omega>0$. In both cases the CTCs are counter-rotating with respect to the
central black hole (or other objects). The implicit assumption here is that the sign of $\Omega$ is the same in the entire exterior spacetime, and so $\text{sgn}(\Omega)=\text{sgn}(\Omega_+)$ for all $r \geqslant r_+$. We wish to discuss an example in which this is \emph{not} the case.

\section{The Curious Case of Kerr-Taub-NUT}\label{III}

\begin{figure}[h!]
\begin{tabular}{rr}
    \includegraphics[width=.5\columnwidth]
      {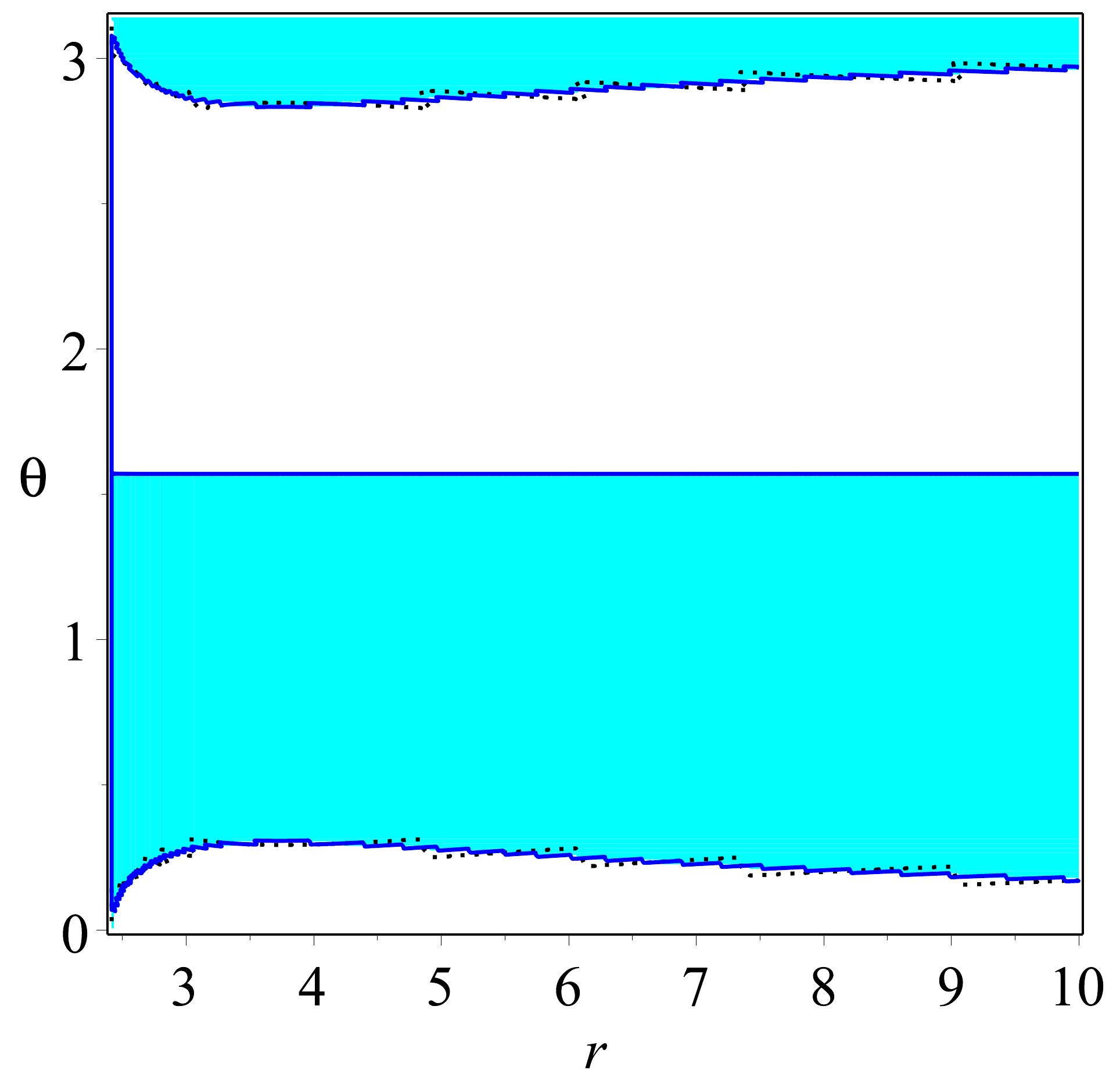}
    \includegraphics[width=.5\columnwidth]
      {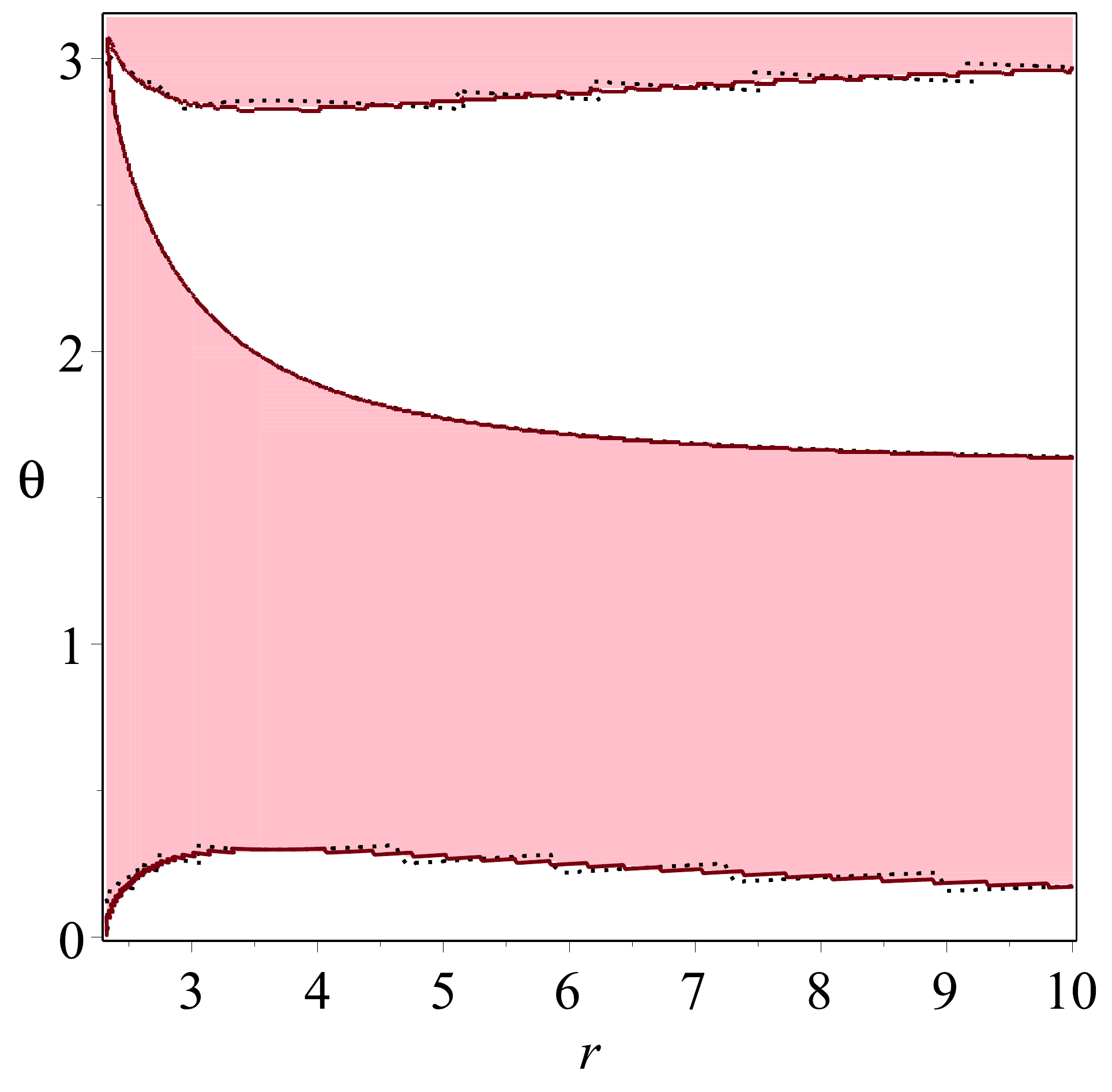}
\end{tabular}
	\caption{The orientation of the angular velocity $\Omega(r,\theta)$ for the Taub-NUT case (left) and the Kerr-Taub-NUT case (right), where $r$ is plotted starting from their respective event horizon. The mass $M$ and the NUT charge $n$ are chosen to be 1, while the rotation parameter $a$ is set to be 0.5 for the Kerr-Taub NUT case. In both cases, CTCs can be found in the shaded region near the south pole, and the unshaded region near the north pole. The shaded regions have opposite rotation with respect to the unshaded region (the solid curves and lines have zero angular velocity). In particular, unlike the Taub-NUT case, the Kerr-Taub-NUT horizon has a well-defined rotating direction. }
	\label{3d}
\end{figure}

The Taub-NUT solution \cite{nut1,nut2,nut3} is well-known to be peculiar even before we introduced a global rotation -- Misner called it ``a counterexample to almost anything'' \cite{any}. It can be described as a ``twisted black hole'', due to the fact that (the exterior of) its northern hemisphere is rotating in a direction opposite to its southern hemisphere \cite{zhang,1610.05757}. Actually, near the poles, just outside the horizon, the angular velocity $\Omega$ is again reversed with respect to the respective hemisphere, as shown in the top panel of Fig.(\ref{3d}).
This spacetime is infested with CTCs near both poles \cite{1610.05757,1610.06135}. It is straightforward to verify that the CTCs counter-rotate with respect to the angular velocity of the \emph{respective} local region. The globally defined \emph{total} angular momentum is however zero since the contributions from the counter-rotating halves cancel out each other.  

\begin{figure}[h!]
		\centering
		\includegraphics[scale=0.32]{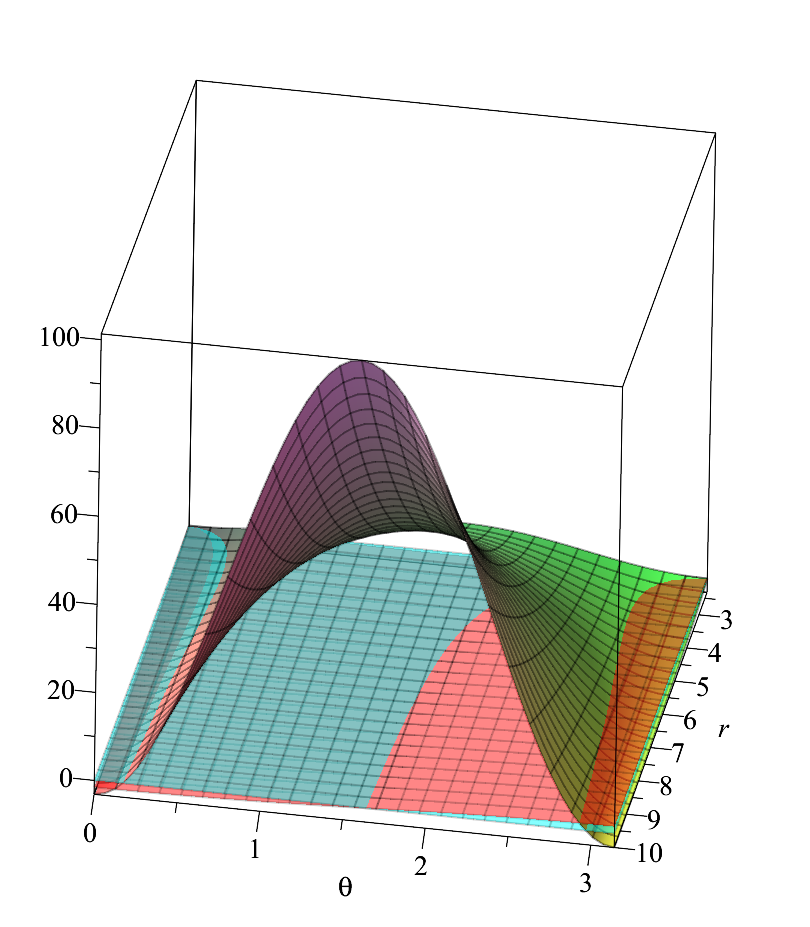}
		\caption{The function $g_{\phi\phi}$ for Kerr-Taub NUT solution becomes negative at the poles (the zero plane is shown in cyan), where there exist CTCs. The function $g_{t\phi}$ is plotted in red: it is positive near the south pole but negative at the north pole, indicating that CTCs are counter-rotating near the south pole but co-rotating at the north pole (both with respect to the horizon). The choice of parameters are $M=n=1, ~a=0.5$.}
		\label{ctc1}
	\end{figure}

We can go on to introduce a global angular momentum and consider instead the Kerr-Taub-NUT solution \cite{DN, Miller}, whose metric tensor in the standard Boyer-Lindquist coordinates give, in particular, the following components:
\begin{equation}\notag
g_{tt}=-\frac{\Delta-a^2 \sin^2\theta}{\Sigma}, ~~
g_{t\phi}=\frac{\Delta\chi-a(\Sigma + a\chi)\sin^2\theta}{\Sigma},
\end{equation}
where 
$\Sigma:=r^2 + (n+a\cos\theta)^2$, $\Delta:=r^2-2Mr+a^2-n^2$, and $\chi:=a\sin^2\theta-2n\cos\theta$.
The parameters $M,a$ and $n$ are respectively the mass, the rotation parameter, and the NUT-charge.
The event horizon $r_+=M+\sqrt{M^2-a^2+n^2}$ now has a well-defined non-vanishing angular velocity, i.e., the entire horizon rotates in a single direction. For $M=1, n=1, a=0.5$, we have the horizon angular velocity $\Omega_+\approx 0.0752$. Sufficiently far outside the horizon each hemisphere still has opposite sense of angular velocity. Again, near the poles the angular velocity changes direction. CTCs are present near both poles (see also \cite{1812.11463}), which are still counter-rotating with respect to the respective local angular velocity. See Fig.(\ref{3d}). Thus the CTCs near the north pole $\theta=0$ are \emph{co-rotating} with respect to the black hole angular velocity, which has the same sign as the total angular momentum \cite{1905.06350}
\begin{equation}
J=\frac{a(r_+^2+a^2+n^2)}{2r_+}.
\end{equation}
We also show the 3-dimensional plot of $g_{\phi\phi}$ and its change of sign in Fig.(\ref{ctc1}).

\section{Discussion: No Obvious Source}
	
The question raised in \cite{0708.2324} is whether CTCs are always counter-rotating with the rotating source. 
We have seen that it is indeed true that CTCs are always counter-rotating with respect to the local angular velocity $\Omega(r,\theta)$, simply due to the defining relation $\Omega=-g_{t\phi}/g_{\phi\phi}$ and the fact that CTC is characterized by $g_{\phi\phi}<0$. The nontrivial question is whether we can understand this in more physical terms. As pointed out in \cite{0708.2324}, the counter-rotation property means that it would be misleading to think of CTCs as being ``caused'' by frame dragging-like effect, which goes in the same direction as the (singly) rotating source. (Actually, even the frame-dragging effect is tricky to be interpreted in general, see below.)

In our work, we see that in a more complicated example in which $\Omega$ can have different signs outside the horizon, it becomes even less clear how CTCs can be caused by the rotation of the source. For the Taub-NUT case ($a=0$), the upper half of the exterior spacetime rotates in an opposite sense from its lower half, hence there is no global angular momentum. The CTCs are counter-rotating with respect to the each rotating half. One could therefore surmise that perhaps the total angular momentum is a red herring, and it is the local rotation that is important. That is, the different halves each act as a source that somehow causes the CTCs to counter-rotate. (Note that the horizon itself does not rotate so the black hole is not quite a ``source''.)
However, in the Kerr-Taub-NUT case, the black hole horizon (the putative source) has a well-defined direction of rotation, yet not all CTCs are counter-rotating with it. In fact, even without considering closed timelike curves, we can ask why $\Omega(r,\theta)$, the angular velocity of a test particle (coming in with zero angular momentum from infinity), can be opposite to that of the angular velocity of the horizon, $\Omega_+$ (which occurs in the unshaded region sandwiched between the shaded regions in the right plot of Fig.(\ref{ctc1}).). That is to say, it is not obviously clear that this ``frame-dragging'' is caused by the rotating black hole dragging spacetime around it.

Perhaps the simplest interpretation is that one should not think of $\Omega$ and the orientation of CTC as being \emph{caused} by the rotation of the black hole too literally. They simply \emph{are}. Of course in the simple case of a Kerr black hole, since $\Omega$ has the same sign as $\Omega_+$ throughout, one can interpret this as a test particle being frame-dragged by the black hole. One can say the same here for particles that are sufficiently close to the horizon of a Kerr-Taub-NUT black hole, but the interpretation ultimately does not work far away from the black hole. 

In the case of CTC of Tipler-van Stockum cylinder, recall also that CTCs occur sufficiently \emph{far} away from the rotating object in the radial direction, not near it. If CTCs are directly ``caused'' by the rotating effect, one typically expects them to occur close to the ``source''. Of course, the Tipler-van Stockum cylinder is infinitely long, and the fields produced by it increases as the distance grows, so in some sense this explains why CTC forms at large distances. 
Nevertheless, the point is that there is no obvious, single simple underlying explanation for the formation and counter-rotation of CTCs, though one could perhaps seek some form of justification case by case.

We should not be too alarmed about the lack of obvious ``cause'' from a ``source'', since such a situation is not new in general relativity \cite{brett} (see also Lecture 17 of \cite{buchdahl}). Recall the interpretation of the mass of a black hole, for example. Take the Schwarzschild black hole for simplicity, which is a \emph{vacuum} solution to Einstein's equation, its mass is then just a property of the nontrivial curvature of the spacetime, and should not be regarded as being ``sourced'' by the singularity, which is afterall, spacelike. In a geometric theory of gravitation like GR, nontrivial spacetime geometries can give rise to surprising behaviors that while mathematically clear, sometimes simply do not admit a good intuitive explanation. This is especially true in our example since the NUT charge has no Newtonian analogue.

\begin{acknowledgments}
YCO thanks the National Natural Science Foundation of China (No.11922508) for funding support. He also thanks Lee-hwa Yeh, from whom he learned about the light cone transformation technique during the summer school ``Physics and Mathematics of General Relativity'', which took place at National Taiwan University in August 2011.
\end{acknowledgments}

\end{document}